\documentclass[
aps,%
preprint,
superscriptaddress
]{revtex4-2}
\usepackage[dvipdfmx]{color}
\usepackage[dvipdfmx]{graphicx}
\usepackage{amsmath,amssymb}
\usepackage{lineno}
\usepackage{gensymb}
\usepackage{textcomp}
\usepackage{xr}
\usepackage[dvipdfmx,colorlinks=true]{hyperref}
\hypersetup{
 linkcolor=blue,
 citecolor=blue,
 urlcolor=blue
 }

\begin{document}
\title{Magnetoelastic anisotropy in Heusler-type Mn$_{2-\delta}$CoGa$_{1+\delta}$ films}%
\author{Takahide Kubota}
\email[]{takahide.kubota@tohoku.ac.jp}
\thanks{author to whom correspondence should be addressed}
\thanks{presently at Department of Advanced Spintronics Medical Engineering, Graduate School of Engineering, Tohoku University, Sendai 980-0845, Japan.}
\affiliation{Institute for Materials Research, Tohoku University, Sendai 980-8577, Japan}
\affiliation{Center for Spintronics Research Network, Tohoku University, Sendai 980-8577, Japan}
\author{Daichi Takano}
\affiliation{Institute for Materials Research, Tohoku University, Sendai 980-8577, Japan}
\author{Yohei Kota}
\affiliation{National Institute of Technology, Fukushima College, Iwaki, Fukushima 970-8034, Japan}
\author{Shaktiranjan Mohanty}
\affiliation{Laboratory for Nanomagnetism and Magnetic Materials (LNMM), School of Physical Sciences, National Institute of Science Education and Research,  An OCC of Homi Bhabha National Institute (HBNI), Jatni 752050, Odisha, India}
\author{Keita Ito}
\affiliation{Institute for Materials Research, Tohoku University, Sendai 980-8577, Japan}
\affiliation{Center for Spintronics Research Network, Tohoku University, Sendai 980-8577, Japan}
\author{Mitsuhiro Matsuki}
\affiliation{Institute for Materials Research, Tohoku University, Sendai 980-8577, Japan}
\author{Masahiro Hayashida}
\affiliation{Institute for Materials Research, Tohoku University, Sendai 980-8577, Japan}
\author{Mingling Sun}
\affiliation{Institute for Materials Research, Tohoku University, Sendai 980-8577, Japan}
\author{Yukiharu Takeda}
\affiliation{Materials Sciences Research Center, Japan Atomic Energy Agency, Hyogo 679–5148, Japan}
\author{Yuji Saitoh}
\affiliation{Materials Sciences Research Center, Japan Atomic Energy Agency, Hyogo 679–5148, Japan}
\author{Subhankar Bedanta}
\affiliation{Laboratory for Nanomagnetism and Magnetic Materials (LNMM), School of Physical Sciences, National Institute of Science Education and Research,  An OCC of Homi Bhabha National Institute (HBNI), Jatni 752050, Odisha, India}
\author{Akio Kimura}
\affiliation{Graduate School of Science, Hiroshima University, Higashi-hiroshima 739–8526, Japan}
\affiliation{Graduate School of Advanced Science and Engineering, Hiroshima University, Higashi-hiroshima 739-8526, Japan}
\author{Koki Takanashi}
\affiliation{Institute for Materials Research, Tohoku University, Sendai 980-8577, Japan}
\affiliation{Center for Spintronics Research Network, Tohoku University, Sendai 980-8577, Japan}
\affiliation{Center for Science and Innovation in Spintronics (Core Research Cluster), Tohoku University, Sendai 980-8577, Japan}
%
%
\begin{abstract}
Perpendicular magnetization is essential for high-density memory application using magnetic materials. High-spin polarization of conduction electrons is also required for realizing large electric signals from spin-dependent transport phenomena. Heusler alloy is a well-known material class showing the half-metallic electronic structure. However, its cubic lattice nature favors in-plane magnetization and thus minimizes the perpendicular magnetic anisotropy (PMA), in general.
This study focuses on an inverse-type Heusler alloy, Mn$_{2-\delta}$CoGa$_{1+\delta}$ (MCG) with a small off-stoichiometry ($\delta$) , which is expected to be a half-metallic material. We observed relatively large uniaxial magnetocrystalline anisotropy constant ($K_\mathrm{u}$) of the order of 10$^5$ J/m$^3$ at room temperature in MCG films with a small tetragonal distortion of a few percent. A positive correlation was confirmed between the $c/a$ ratio of lattice constants and $K_\mathrm{u}$. Imaging of magnetic domains using Kerr microscopy clearly demonstrated a change in the domain patterns along with $K_\mathrm{u}$.
X-ray magnetic circular dichroism (XMCD) was employed using synchrotron radiation soft x-ray beam to get insight into the origin for PMA. Negligible angular variation of orbital magnetic moment ($\Delta m_\mathrm{orb}$) evaluated using the XMCD spectra suggested a minor role of the so-called Bruno's term to $K_\mathrm{u}$.
Our first principles calculation reasonably explained the small $\Delta m_\mathrm{orb}$ and the positive correlation between the $c/a$ ratio and $K_\mathrm{u}$. The origin of the magnetocrystalline anisotropy was discussed based on the second-order perturbation theory in terms of the spin--orbit coupling, claiming that the mixing of the occupied $\uparrow$- and the unoccupied $\downarrow$-spin states is responsible for the PMA of the MCG films.	
\end{abstract}
\date{\today}%
%
\maketitle
\section{Introduction}
Heusler alloys have been attracting interest because of their  various physical properties with selection of a plenty of compositions~\cite{Felser2016}. Especially, in spintronics research field, half-metallic electronic structure, providing fully spin-polarized conduction electrons, has motivated spin-dependent transport studies such as magnetoresistance effects~\cite{Tanaka1997,Caballero1998}.
Among several kinds of Heusler alloys, Co-based Heusler alloys, \textit{e.g.} Co$_2$\textit{YZ}, \textit{Y}: a transition metal, \textit{Z}: a 12 or 13 group element, are representative materials that have been predicted to show the half-metallicity~\cite{Kubler1983,Ishida1995,Galanakis2002,Miura2004,Kandpal2006}.
The half-metallicity was experimentally deduced from the large magnetoresistance effects,~\cite{Inomata2003,Sakuraba2005,Schmalhorst2008,Kubota2009,Yamamoto2010,Liu2015,Sakuraba2012,Kubota2019cpp} the X-ray absorption spectroscopy~\cite{Miyamoto2003,Telling2006,Yoshikawa2020,Kono2020} and photoelectron spectroscopy~\cite{Kolbe2012,Jourdan2014}. Recent bulk-sensitive angle-resolved photoelectron spectroscopy experiments of Co$_{2}$MnGe with soft X-ray synchrotron radiation beam uncovered three-dimensional energy dispersions, which are in good agreement with the theoretical band structures~\cite{Kono2020}.
In contrast to Co-based Heusler alloys, Mn-based Heusler alloys, Mn$_2$\textit{YZ} gained renewed interests as a new class of half-metallic Heusler alloys~\cite{Ishida1984,Galanakis2002}. The major advantage of the Mn-based Heusler alloy is relatively small saturation magnetization ($M_\mathrm{s}$) which reduces critical current density for current induced magnetization switching phenomena based on spin-transfer-torque (STT)~\cite{Slonczewski1996}.
In addition, some of the Mn-based Heusler alloys with the inverse-Heusler structure (X$_\mathrm{A}$ phase) were predicted to possess novel electronic structures, \textit{e.g.} half-metallic spin-gapless semiconductor~\cite{Ouardi2013,Galanakis2014,Xin2017}.
Among the Mn-based Heusler alloys, Mn$_2$CoGa is highlighted in this paper.
Mn$_2$CoGa is a theoretically predicted half-metallic inverse-type Heusler alloy~\cite{Alijani2011,Klaer2011,Chadov2012}.
The composition dependent magnetocrystalline anisotropy is an interesting feature in the Mn$_2$CoGa and the related compositions: Tetragonal distortion of the lattice occurs due to the band Jahn-Teller effect in Mn$_{3-x}$Co$_x$Ga alloys with \textit{x} below 0.5, and relatively large magnetocrystalline anisotropy of the order of 10$^6$ J/m$^3$ has been  observed~\cite{Alijani2011,Ouardi2012,Kubota2013}.
The tetragonal distortion destroys the half-metallic energy gap, while the conduction electron maintains relatively high spin polarization due to the difference in the electron mobilities between the majority and minority spin channels~\cite{Chadov2012}.
Recent studies also reported another interesting property of a Mn$_2$CoGa film showing perpendicular magnetic anisotropy (PMA) originating from the magnetocrystalline anisotropy in a nearly cubic lattice~\cite{Kubota2014}.
In the previous study, Mn$_{2-\delta}$CoGa$_{1+\delta}$ (MCG) films ($\delta\sim$ 0.1) grown onto MgO (001) single crystal substrates exhibited perpendicular magnetization and in-plane magnetization with and without a Cr buffer layer underneath, respectively. Such a drastic change can be ascribed to a small difference in the lattice strain, $c/a$ ratios which were 1.00 and 1.01 for the films without and with the Cr buffer layer, respectively.
The previous results suggested that the small lattice strain induced magnetocrystalline anisotropy in MCG in the cubic phase, \textit{i.e.}, the magnetoelastic anisotropy was induced, which is totally different from the interface PMA in Co-based full-Heusler alloys containing Fe, for which PMA energy of the order of 10$^5$ J/m$^3$ was reported in ultra-thin, $\sim$ 1 nm, films layered with MgO~\cite{Wang2010,Wen2011,Kamada2014,Kubota2016}.
The PMA in highly spin polarized materials is essential for spintronic devices, such as STT-type magnetoresistive random access memories. Concerning the MCG, however, the half-metallicity in the strained lattice as well as the origin of  the magnetocrystalline anisotropy remains unsolved thus far.

In this study, a set of MCG epitaxially grown films showing different PMA and $c/a$ ratios were prepared, and correlations between the magnetic properties, magnetic domain images, and electronic structures were discussed based on experimental results and theoretically calculated electronic structures.
To figure out the chemical order and electronic structures of film samples, synchrotron radiation soft x-ray absorption spectra (XAS) and soft x-ray magnetic circular dichroism (XMCD) were also investigated.
There were many reports with XAS and XMCD results on Heusler alloys containing Mn atoms~\cite{Miyamoto2003,Telling2006,Nagai2018} including Mn$_2$CoGa~\cite{Klaer2011,Meinert2011,Ouardi2015}. In the previous studies, the degree of the electron localization was discussed using the XAS and XMCD spectra, which are also addressed in this study.

\section{Experimental Procedures}
\label{sec:exp}
All samples were deposited onto MgO(001) single-crystal substrates by using an ultrahigh-vacuum magnetron sputtering machine with a base pressure below $3 \times 10^{-7}$ Pa. The MCG layers were prepared by co-sputtering technique using a Mn$_{55}$Ga$_{45}$ target and a Co target. The film composition of the MCG layer was Mn$_{42.6}$Co$_{24.3}$Ga$_{33.1}$ (at.\%) corresponding to Mn$_{1.7}$CoGa$_{1.3}$ in the atomic ratio, which was characterized by an electron probe micro analyzer with a standard sample for the quantitative analysis.
We employed two materials for the buffer layer dependence study: Ag and Cr. The stacking structures of samples and the name for the sample series were as follows:
Series A (Ag buffer): MgO sub. /Cr (20 nm) /Ag (40 nm) / MCG ($t$) /Ta (3 nm), and
series B (Cr buffer): MgO sub. /Cr (20 nm) / MCG ($t$) /Ta (3 nm).
The thicknesses ($t$) of the MCG layer were 5, 10, 20 and 30 nm. Before installing the samples inside the vacuum chamber, the substrates were cleaned using acetone and ethyl alcohol, and after introducing them inside the chamber, the substrates were annealed at 700 {\degree}C to clean the surface. All layers were deposited at room temperature, and \textit{in situ} post-annealing was carried out at 700 {\degree}C (500 {\degree}C) for the Cr (MCG) layer. The morphology of the sample surfaces was observed by using an atomic force microscope (AFM) after depositing a Ta capping layer. The crystal structures and magnetic properties were characterized by x-ray diffraction (XRD), and using a vibrating sample magnetometer (VSM), respectively, at room temperature.

The hysteresis loop and domain images were recorded simultaneously by magneto-optical Kerr effect (MOKE) microscopy manufactured by Evico Magnetics Ltd. Germany. All the samples were measured in polar MOKE (P-MOKE) geometries in which the sample surface was perpendicular to the applied magnetic field.

The XAS and XMCD measurements were carried out on $L_{2,3}$-edges of Mn and Co atoms by a total electron-yield method at the twin helical undulator beamline BL23SU of SPring--8~\cite{Saitoh2012}.
External magnetic fields ($H$) of $\pm$3 T and $\pm$8 T were applied for measurements at out-of-plane field ($\theta$: 0\degree) and pseudo in-plane field ($\theta$: 54.7\degree), respectively. The measurement temperature was set at 300 K.

\section{Results and Discussion}
\subsection{Surface Morphology and Crystal Structure}
\begin{figure}
\includegraphics[clip,scale=1.0]{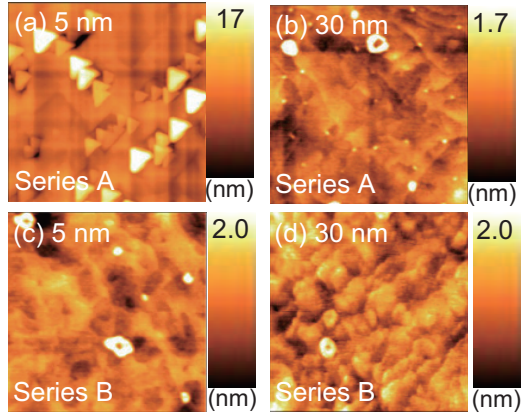}
\caption{Atomic force microscope images of the sample surfaces. Series A: (a) $t$: 5, and (b) 30 nm. Series B: (c) $t$: 5, and (d) 30 nm. Scan-area is $1 \times 1 \mu \mathrm{m}^2$ for all the images.}
\label{fig:AFM}
\end{figure}
Surface morphologies were observed using AFM and the results are shown in figures~\ref{fig:AFM}(a)--\ref{fig:AFM}(d) for the series A and B with $t$ of 5 and 30 nm. For the 5-nm-thick samples, several islands with a height of several nanometers are observed in series A (Fig.~\ref{fig:AFM}(a)), while relatively flat surface is observed in series B (Fig.~\ref{fig:AFM}(c)). Regarding the 30-nm-thick samples, the surfaces are relatively flat for both series A and B.
\begin{figure}
\includegraphics[clip,scale=1.0]{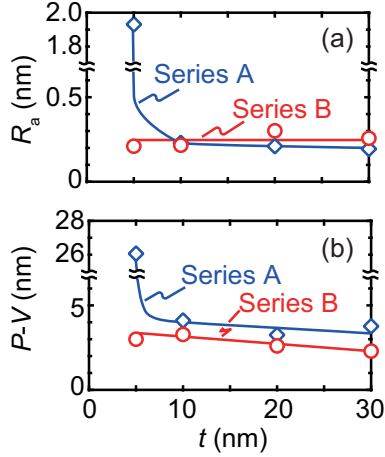}
\caption{Summary of (a) average surface roughness ($R_\mathrm{a}$) and (b) peak-to-valley height (\textit{P}-\textit{V}) values as a function of $t$.}
\label{fig:Ra}
\end{figure}
The thickness dependence of the average surface roughness ($R_\mathrm{a}$) and peak-to-valley height (\textit{P}-\textit{V}) are summarized in figures.~\ref{fig:Ra}(a) and~\ref{fig:Ra}(b), respectively. The values of $R_\mathrm{a}$ and \textit{P}-\textit{V} drastically drop at $t = 10$ nm for series A, while, those values show no dependence on $t$ and relatively small values for series B. These results suggest a difference in the growth mode of the MCG layers depending on the buffer material: For the series A using Ag buffer layer, the Volmer-Weber (VW) mode, which describes a three-dimensional growth, is considered for the initial growth of the MCG layer. It then possibly changes to the Frank-van-der-Merwe (FM) mode for a layer-by-layer growth mode with increasing $t$.
For the series B using Cr buffer layer, the film was grown possibly in the FM mode throughout the whole thicknesses.
The lattice spacings of both buffer materials are smaller than that of a bulk Mn$_2$CoGa~\cite{Alijani2011}, and the lattice mismatches are approximately 1.6\% and 1.4\% to the Cr and Ag buffer layers, respectively, which are of the same order. A reason for the difference in the growth modes between the two sample series can be qualitatively understood by considering surface energies for the buffer$/$MCG layer system in the following way:
According to the quasi-equilibrium description, a sum of surface energies ($\sigma_\mathrm{sum}$) can be written as $\sigma_\mathrm{sum} = \sigma_\mathrm{f} - \sigma_\mathrm{b} + \sigma_\mathrm{i}$, where $\sigma_\mathrm{f}$, $\sigma_\mathrm{b}$, and $\sigma_\mathrm{i}$ are surface energies of a film, a buffer layer, and the interface energy between the film and the buffer layer, respectively~\cite{Bauer1982,Kief1993}.
Here, strain energies for the film and the buffer layer are neglected because the lattice strain is small ($\sim$ a few \%) in the present samples. We may also neglect the difference between $\sigma_\mathrm{i}$ values for the cases of series A (Ag$/$MCG interface) and series B (Cr$/$MCG interface), because $\sigma_\mathrm{i}$ depends on the lattice misfits between the layers which are comparable to each other for the present sample series. In addition, contribution of $\sigma_\mathrm{i}$ to the value of $\sigma_\mathrm{sum}$ could be very small because the potential energy for the interface energy exhibits exponential decrease when the misfit is close to zero~\cite{Merwe1963a,Merwe1963b,Fletcher1964}.
Based on the points stated above, $\sigma_\mathrm{f}$ and $\sigma_\mathrm{b}$ can be dominant parameters affecting the sign of $\sigma_\mathrm{sum} \sim \sigma_\mathrm{f} - \sigma_\mathrm{b}$, \textit{i.e.} VW (FM) mode is expected for $\sigma_\mathrm{sum} > 0$ ($ < 0$) for which $\sigma_\mathrm{f}$ is larger (smaller) than $\sigma_\mathrm{b}$.
\begin{center}
\begin{table}
\caption{
Reported surface energies in literature.
}
\label{t:surface}
\begin{ruledtabular}
\begin{tabular}{ c  c c c}
 & \multicolumn{3}{c}{Surface energy (J/m$^2$)}\\
\hline
Refs. & \onlinecite{Mezey1982}
\footnote{Thermodynamic calculation for poly-crystalline system.}
& \onlinecite{Guisbiers2018}
\footnote{Experimental value for poly-crystalline samples.}
& \onlinecite{Vitos1998}
\footnote{Calculated values for (001) surface based on full-charge density linear muffing-tin orbitals method.}\\
\hline
Cr& 2.056 & 2.300 &  3.979 \\
Ag & 1.302 & 1.250 & 1.200\\
\end{tabular}
\end{ruledtabular}
\end{table}
\end{center}
Reported surface energies for poly-crystalline and (001) surface of the cubic systems are summarized in Table~\ref{t:surface}~\cite{Mezey1982,Guisbiers2018,Vitos1998}.
Although the surface energy of MCG is unknown here, we may qualitatively understand the different growth modes considering the surface energies of Cr and Ag as $\sigma_\mathrm{b}$, \textit{i.e.}, the surface energies of Cr are larger than those of Ag for all the cases shown in Table~\ref{t:surface}. Thus, the $\sigma_\mathrm{sum}$ can be relatively small for the Cr buffer sample compared to that using the Ag buffer sample, which possibly resulted in better coverage of the MCG layer deposited on the Cr buffer than that on the Ag buffer.
In addition to the possibility of the different growth modes explained above, different situations for the surface reconfiguration caused by the post-annealing can be another factor: \textit{i.e.}, the MCG layer forms islands (the layer structure) on the Ag (Cr) buffer after the annealing because of the relatively small (large) $\sigma_\mathrm{b}$, especially for the relatively thin 5-nm-thick samples.
For $t \geq$ 10 nm, the  good coverage was likely realized because of the thick (relatively large volume) of the MCG layer for the as-deposited stage and surface reconfiguration caused by the post-annealing.

\begin{figure*}
\includegraphics[clip,scale=1.0]{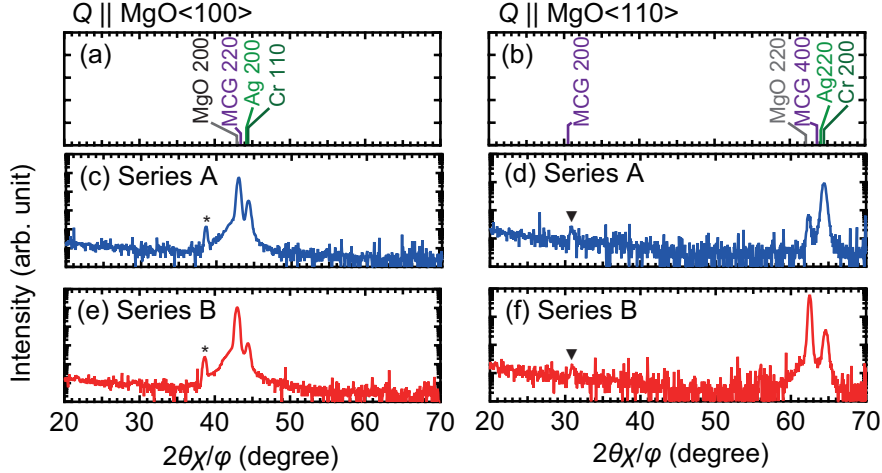}
\caption{In-plane XRD profiles of 5-nm-thick samples. (a, b) Diffraction positions for MCG, Ag, Cr layers, and MgO substrates. (c, d) Diffraction profiles of series A, and (e, f) those of series B. Diffraction vectors ($\vec{Q}$) are parallel to MgO$<$100$>$ and MgO$<$110$>$ for (c, e) and (d, f), respectively. Weak peaks around 67\textdegree -- 69\textdegree~in (c) and (e) are probably from the Ta capping layers.}
\label{fig:xrd}
\end{figure*}
The out-of-plane and in-plane XRD measurements were carried out to investigate the crystal structure.
The XRD profiles of 5-nm-thick samples are shown in figure~\ref{fig:xrd} for series A (Figs.~\ref{fig:xrd}(c) and~\ref{fig:xrd}(d)) and B (Figs.~\ref{fig:xrd}(e) and~\ref{fig:xrd}(f)). From the MCG layer, a series of $h~0~0$ ($h =$ 2 or 4) diffractions are observed for the diffraction profile with $\vec{Q} || $MgO[110], and only 2~2~0 diffractions are observed for $\vec{Q} ||$MgO[100] where $\vec{Q}$ is a diffraction vector. Features of these diffraction profiles were similar for all other samples indicating the epitaxial growth with a relationship of MgO(001)[100] $|$ MCG(001)[110].
\begin{figure}
\includegraphics[clip,scale=1.0]{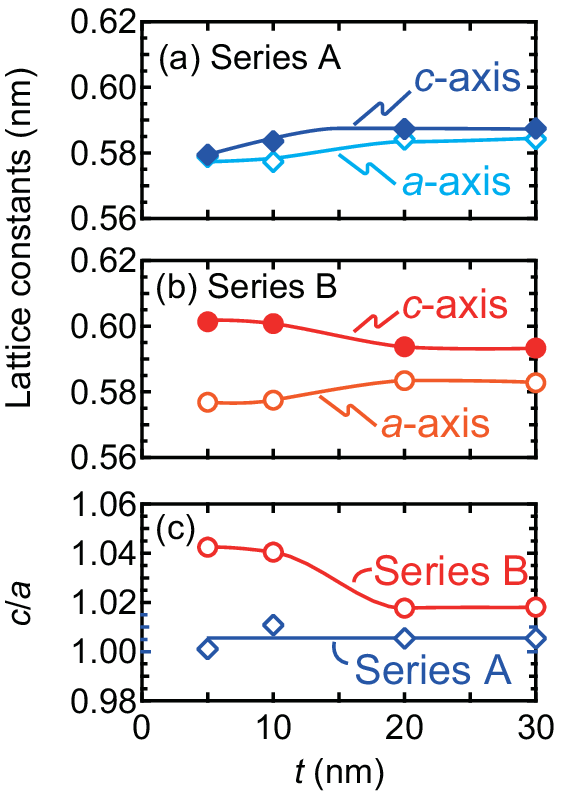}
\caption{Thickness ($t$) dependence of lattice constants of the samples: (a) series A on Ag buffers, (b) series B on Cr buffers, and (c) the $t$ dependence of c/a ratios for series A and B. The error bars range within the size of data points}
\label{fig:lattice}
\end{figure}
Figures~\ref{fig:lattice}(a),~\ref{fig:lattice}(b), and \ref{fig:lattice}(c) show the $t$ dependence of the out-of-plane lattice constant ($c$), the in-plane lattice constants ($a$), and the $c/a$ ratio, respectively.
For the series A, the $c/a$ ratio shows nearly no dependence on $t$, on the other hand for the series B, although the $c/a$ ratio is 1.04 at $t = 5$ nm and relaxes with increasing $t$, the tetragonal distortion is maintained at a relatively thick $t$ of 30 nm for which the $c/a$ ratio is 1.02.
The different lattice strain possibly originates from the difference in the growth modes of the MCG layers: For the series A, as it was found in the observation of morphology using the AFM images, the strain inside the lattice could be relaxed because of the three-dimensional growth mode in the initial stage. For the series B, on the other hand, the layer-by-layer growth in the initial stage resulted in the restriction of the MCG lattice to the in-plane direction, which remained unrelaxed even at the relatively large thickness.

\subsection{Magnetic Properties}
\label{sec:mag}
\begin{figure}
\includegraphics[clip,scale=1.0]{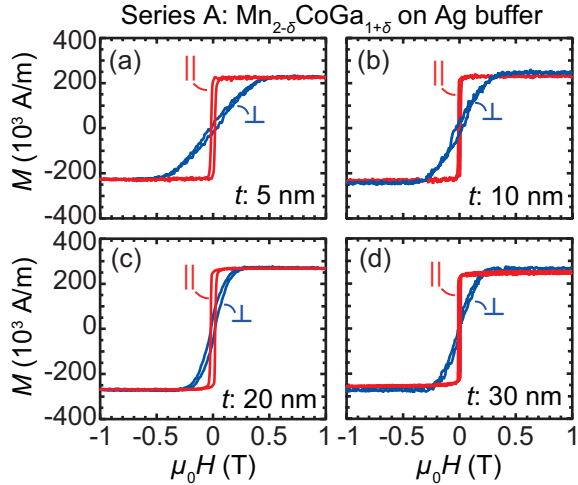}
\caption{Magnetization curves for series A samples. Marks $\perp$ and $||$ represent the out-of-plane and in-plane directions, respectively, for applied magnetic fields. The layer thicknesses ($t$) are (a) 5 (b) 10, (c) 20, and (d) 30nm. The measurements were carried out at room temperature.}
\label{fig:MH_A}
\end{figure}%
\begin{figure}
\includegraphics[clip,scale=1.0]{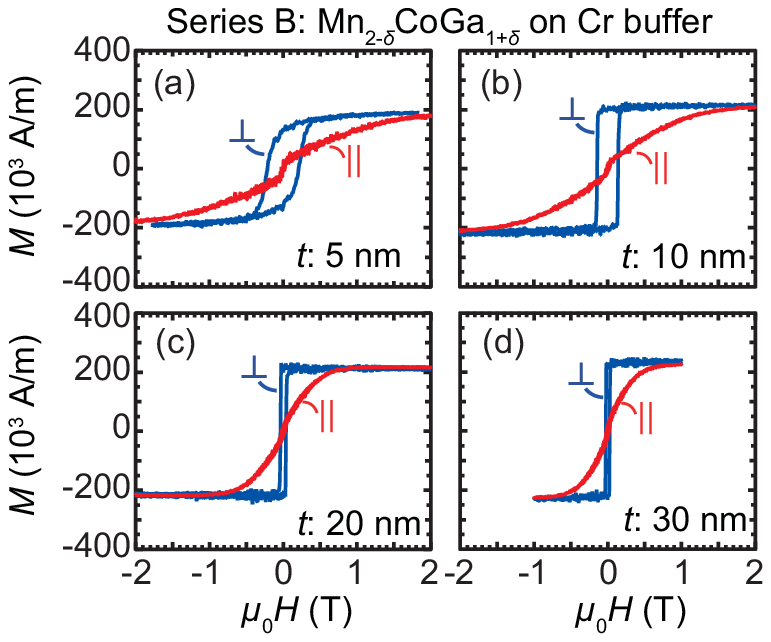}
\caption{Magnetization curves for series B samples. Marks $\perp$ and $||$ represent the out-of-plane and in-plane directions, respectively, for applied magnetic fields. The layer thicknesses ($t$) are (a) 5 (b) 10, (c) 20, and (d) 30nm. The measurements were carried out at room temperature.}
\label{fig:MH_B}
\end{figure}
Figures~\ref{fig:MH_A} and~\ref{fig:MH_B} show magnetization curves of all samples for the series A and B, respectively. In-plane magnetization is observed for the series A, and perpendicular magnetization is observed for the series B. The perpendicular magnetization in series B is consistent with a previous study on a Mn$_{1.8}$Co$_{1.2}$Ga$_{1.0}$ film~\cite{Kubota2014}.
These features for the easy magnetization directions were the same for all other layer thicknesses down to 5 nm in each sample series.
\begin{figure}
\includegraphics[clip,scale=1.0]{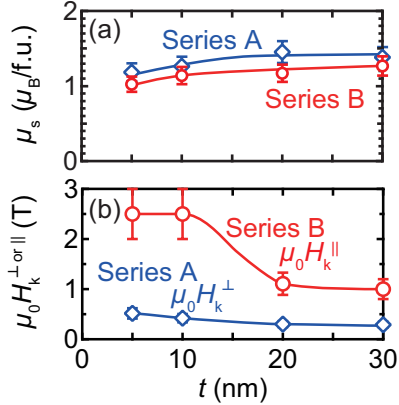}
\caption{(a) Saturation magnetic moment ($\mu_\mathrm{s}$) per a formula unit (f.u.)
and (b) saturation field ($H_\mathrm{k}^{\perp \mathrm{(or}~||\mathrm{)}}$) as a function of $t$, where  $H_\mathrm{k}^{\perp \mathrm{(or}~||\mathrm{)}}$ represents anisotropy field for out-of-plane (in-plane) direction. The measurements were carried out at room temperature.}
\label{fig:mag}
\end{figure}
Figure~\ref{fig:mag}(a) shows the $t$ dependence of 
saturation magnetic moments ($\mu_\mathrm{s}$) which were calculated using $M_\mathrm{s}$ values obtained from the magnetization curves (Figs.~\ref{fig:MH_A} and \ref{fig:MH_B})
and the lattice constants (Fig. \ref{fig:lattice}). The $\mu_\mathrm{s}$ shows nearly no dependence on $t$, which suggests a similar degree of chemical order among the samples even for the relatively low thickness region.
The magnitudes of $\mu_\mathrm{s}$ range from 1.1 to 1.4 $\mu_\mathrm{B}$ per a formula unit (f.u.), which are relatively small compared to other ferromagnetic Heusler alloys showing the moment of the order of 4 -- 6 $\mu_\mathrm{B}$ because of the ferrimagnetic alignment of the atomic moments.
Although the experimental values are smaller than theoretical values of approximately 2 $\mu_\mathrm{B}$ for Mn$_2$CoGa in literature~\cite{Umetsu2018} as well as in our calculation (see Sec.~\ref{sec:calc}), those are of the same order of other calculated values of about 1 $\mu_{B}$ in which the off-stoichiometry and the disorder effects were considered~\footnote{The calculation details for the magnetic moments with the offstoichiometry and the disorder are described in Supplementary Material.}.
Comparing the two sample series, the samples in series A exhibit slightly larger $\mu_\mathrm{s}$ than those in series B. The difference is possibly because of a small difference of chemical order for each sample series\cite{Kubota2009a}.
As another possibility, the difference in composition distribution probably caused by a difference in a small amount of interdiffusion around the Ag (or Cr) / MCG interface may be considered\cite{Xu2019}.
Figure~\ref{fig:mag}(b) shows the $t$ dependence of saturation field for hard magnetization directions which are the out-of-plane field ($H_\mathrm{k}^\perp$) and the in-plane field ($H_\mathrm{k}^{||}$) for series A and B, respectively. The $H_\mathrm{k}^{\perp}$ monotonically decreases as $t$ increases for the series A, while $H_\mathrm{k}^{||}$ shows a drop around the $t$ of greater than 10 nm for the series B.
\begin{figure}
\includegraphics[clip,scale=1.0]{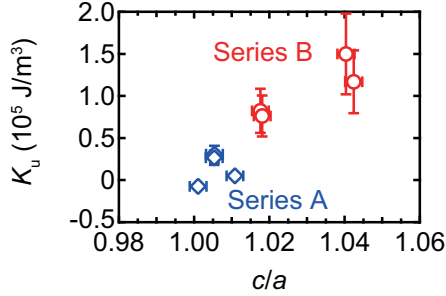}
\caption{The $c/a$ ratio dependence of uniaxial magnetocrystalline anisotropy constant ($K_\mathrm{u}$).}
\label{fig:Ku}
\end{figure}
The uniaxial magnetocrystalline anisotropy constant ($K_\mathrm{u}$) of MCG films was evaluated from the magnetization curves using the following formula; $K_\mathrm{u} = K_\mathrm{u}^\mathrm{eff} + (\mu_0/2)M_\mathrm{s}^2$, where $(\mu_0/2)M_\mathrm{s}^2$ is the shape anisotropy energy, and $K_\mathrm{u}^\mathrm{eff}$ was derived from the area enclosed by the out-of-plane and in-plane magnetization curves.
A positive value of $K_\mathrm{u}^\mathrm{eff}$ represents the perpendicular magnetization in this study.
The values of $K_\mathrm{u}$ are summarized as a function of the $c/a$ ratios in figure~\ref{fig:Ku}. Here, the $K_\mathrm{u}$ exhibits positive correlation with the $c/a$ ratio indicating that it is caused by the magnetoelastic effect. The maximum value of $K_\mathrm{u}$ was 1.6 $\times$ 10$^5$ J/m$^3$ at the $c/a$ of 1.04. The $c/a$ dependent $K_\mathrm{u}$ is theoretically discussed in Sec.~\ref{sec:calc}. Although the strain could be maximum around the interface region and be relaxed around the surface region in the real samples, the following discussion is based on the uniform strain for the simplicity.

\subsection{Domain Observations}
\begin{figure}
\includegraphics[clip,scale=1.0]{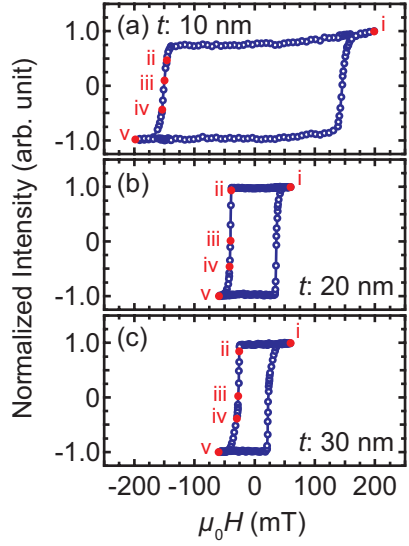}
\caption{The hysteresis loops for $t =$ (a) 10, (b) 20, and (c) 30 nm in series B measured with Kerr microscopy in polar mode. The red dots and numbers i -- v on the hysteresis loops corresponds to the points for which domain images are shown in the next figure \textit{i.e.}, Fig.~\ref{fig:MOKEimages}.}
\label{fig:MOKEloop}
\end{figure}
\begin{figure*}
\includegraphics[clip,scale=1.0]{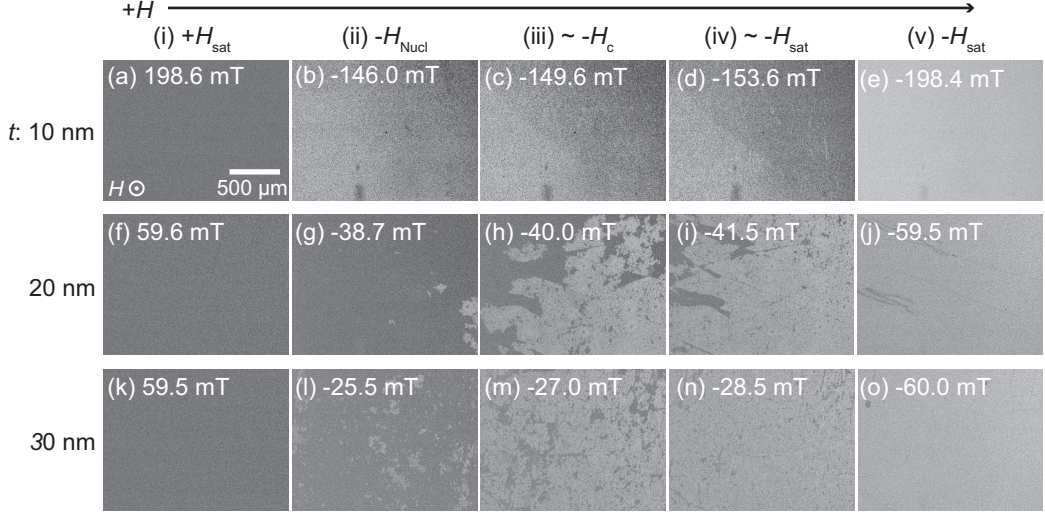}
\caption{The domain images for (a -- e) $t$ = 10, (f -- j) 20, and (k -- o) 30 nm in series B, whose hysteresis loops are shown in Figs.~\ref{fig:MOKEloop}(a),~\ref{fig:MOKEloop}(b), and~\ref{fig:MOKEloop}(c), respectively. The field points (i -- v) which are shown above the panels are annotated  in Fig.~\ref{fig:MOKEloop}. The exact field values are shown in each panel. The scale bar and field direction are shown in (a), which is valid for all the images. }
\label{fig:MOKEimages}
\end{figure*}

The values of $K_\mathrm{u}$ exhibit clear dependence on $t$, especially in the samples of series B, for which the diﬀerence in magnetic domain structures was also observed using Kerr microscopy~\cite{Belhi2014,Mallick2018}.
Figure~\ref{fig:MOKEloop} shows hysteresis loops for the samples of series B with $t =$ 10 (Fig.~\ref{fig:MOKEloop}(a)), 20 (Fig.~\ref{fig:MOKEloop}(b)) and 30 nm (Fig.~\ref{fig:MOKEloop}(c)) measured by MOKE in polar mode. The square like shapes of the hysteresis indicate that the magnetization reversal occurs via domain nucleation and domain wall motion for all the samples~\cite{Belhi2014,Hubert1998}.
To clarify the change of domain pattern, five points were chosen for MOKE microscopy observation: (i) a saturation point at a positive field ($+H_\mathrm{sat}$), (ii) a point around nucleation at a negative field ($-H_\mathrm{nucl}$), (iii) a point around coercivity at a negative field ($-H_\mathrm{c}$), (iv) a point near saturation in a negative field ($\sim -H_\mathrm{sat}$), and (v) a saturation point for the negative field ($-H_\mathrm{sat}$). The numbers, i -- v are also annotated in each panel of Fig.~\ref{fig:MOKEloop}, and the respective field values are shown in each panel of MOKE images in Fig.~\ref{fig:MOKEimages}.
The domain images observed by MOKE microscopy in polar mode of three samples are shown in Fig.~\ref{fig:MOKEimages}. Note that the domain image was not studied for $t$ = 5 nm because of large $H_\mathrm{c}$ ($\mu_0H_\mathrm{c}\sim$ 210 mT) which was out of the range of the MOKE setup.
For the 10-nm-thick sample, relatively large size of domains appears, which contain a 180\degree~domain wall and are visible in Figs.~\ref{fig:MOKEimages}(b) to~\ref{fig:MOKEimages}(d).
The domain pattern changes for $t$ = 20 nm, in which large patch-like domains are observed (Figs.~\ref{fig:MOKEimages}(g) --~\ref{fig:MOKEimages}(i)). With increasing $t$ to 30 nm, the domain size becomes small (Figs.~\ref{fig:MOKEimages}(l) --~\ref{fig:MOKEimages}(n)).
It is observed here that the domain size is observed to be comparatively large for the sample with large anisotropy energy. 
The surface roughness likely causes the magnetic domain pinning\cite{Bruno1990}.

\subsection{X-ray Magnetic Circular Dichroism}
\label{sec:XMCD}
\begin{figure*}
\includegraphics[clip,scale=1.0]{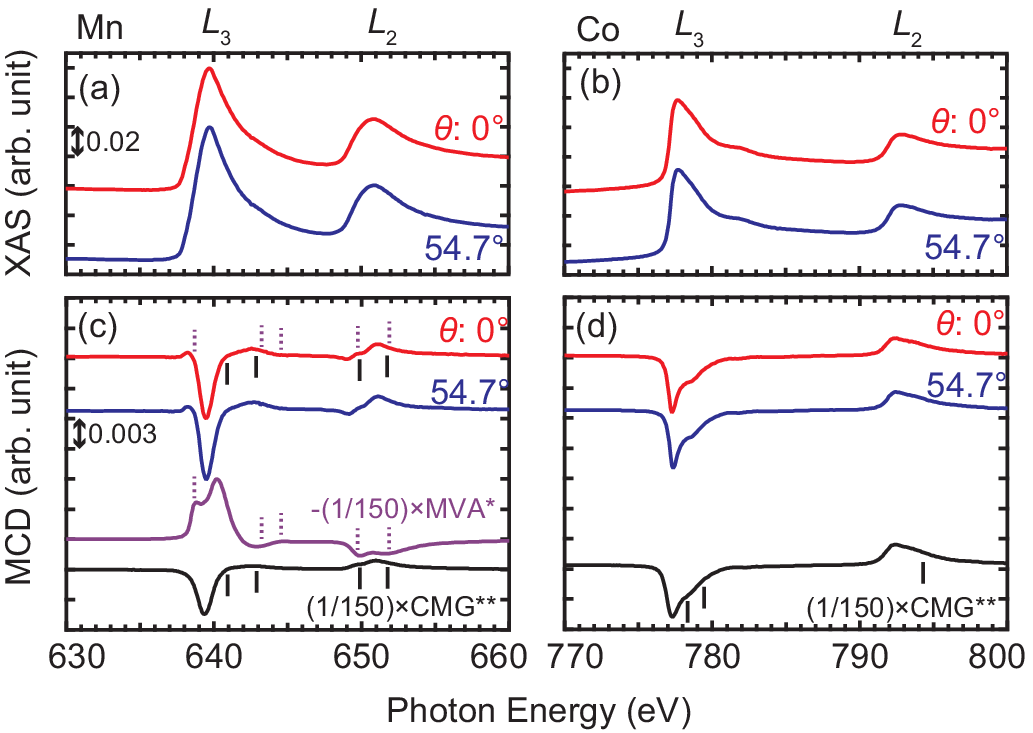}
\caption{
(a, b) Normalized soft x-ray absorption spectra (XAS) and (c, d) soft x-ray magnetic circular dichroism (XMCD) spectra for the series B sample around the (a, c) Mn $L_{2,3}$ and (b, d) Co $L_{2,3}$ absorption edges.  Red lines and blue lines represent spectra taken at $\theta = 0\degree$ and 54.7\degree, respectively, where  $\theta = 0\degree$ is the perpendicular to the film plane direction.
MVA and CMG are reference spectra of bulk Mn$_2$VAl~\cite{Nagai2018} and Co$_2$MnGa~\cite{Yoshikawa2020} samples, respectively.
Positions for the shoulders in XMCD spectra are marked by dotted lines and solid lines for those in MVA and CMG, respectively, and the corresponding structures are also marked in the spectra of series B sample using the same markers.
}
\label{fig:MCD}
\end{figure*}
\begin{figure}
\includegraphics[clip,scale=0.8]{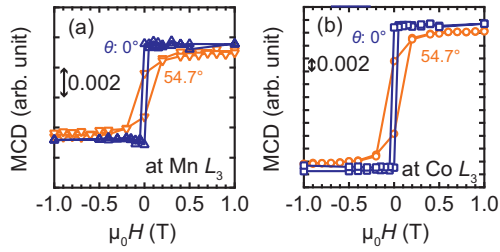}
\caption{
Element specific hysteresis loops measured at $L_3$ absorption edges of (a) Mn and (b) Co atoms. Angles, $\theta$ shown in each pannel represent magnetic fied direction respect to the sample normal.}
\label{fig:ESMHs}
\end{figure}
\begin{figure}
\includegraphics[clip,scale=0.8]{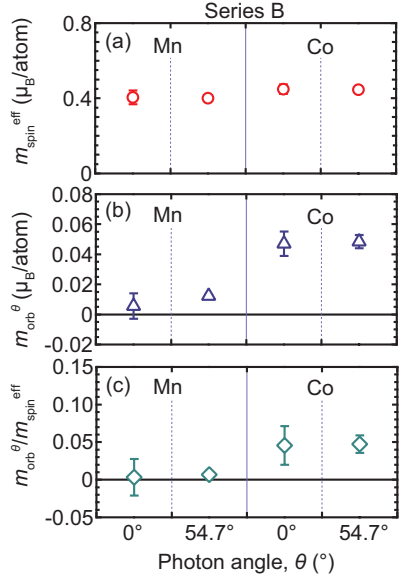}
\caption{
Summary of (a) the effective spin-moment ($m_\mathrm{spin}^\mathrm{eff}$) and (b) orbital moment ($m_\mathrm{orb}^\mathrm{\theta}$) per a atom evaluated using magneto-optical sum rule. (c) $m_\mathrm{orb}^\mathrm{\theta} / m_\mathrm{spin}^\mathrm{eff}$ values. Photon angles, $\theta$ =  0\degree\ and 54.7\degree\ are defined as the out-of-plane and pseudo-in-plane directions, respectively. All data points are for the series B sample.
}
\label{fig:sumrule}
\end{figure}

Element specific magnetic moment of the 30-nm-thick sample showing perpendicular magnetization (series B) was characterized using synchrotron radiation soft x-ray at BL23SU of SPring-8~\cite{Saitoh2012}. The XAS ($= \mu^- + \mu^+$) and XMCD ($= \mu^- - \mu^+$) spectra are shown in figure~\ref{fig:MCD} for Mn $L_{2,3}$ and Co $L_{2,3}$ absorption edges measured at $\theta$ of 0\degree\ and 54.7\degree, where $\mu^{+(-)}$ represents the soft x-ray absorption for the positive (negative) helicity.
The XAS and XMCD spectra of previously studied bulk Mn$_2$VAl~\cite{Nagai2018} and Co$_2$MnGa~\cite{Yoshikawa2020} samples are also shown in Fig.~\ref{fig:MCD} to discuss the origin of structures observed in the spectra for the series B sample, which is to be described in the latter part of this section.
Element specific hysteresis loops measured at the $L_3$ absorption edges of Mn and Co atoms are shown in figure~\ref{fig:ESMHs}. MCD signals for Mn and Co atoms simultaneously switch around the coercive field. For $\theta = 0$, the coercive field is consistent with that observed in VSM measurements shown in Figs.~\ref{fig:MH_A} and ~\ref{fig:MH_B}. These results indicate strong ferromagnetic coupling between the Co moments and the net moment for Mn atoms. 
To quantitatively discuss the angular dependence of XMCD, orbital- ($m_\mathrm{orb}^\theta$) and effective spin-moments ($m_\mathrm{spin}^\mathrm{eff}$) per an atom were evaluated using magneto-optical sum rules~\cite{Thole1992,Carra1993,Stohr1995,van_der_Laan2014,Koide2001,Edmonds2015}  which state the moments using the following equations;
\begin{equation}
\label{eq:morb}
m_\mathrm{orb}^\theta = -\frac{4q}{3r} n_\mathrm{h},
\end{equation}
\begin{equation}
\label{eq:mspin}
m_\mathrm{spin}^\mathrm{eff} = m_\mathrm{spin} + 7m_T^\theta = - \frac{6p - 4q}{r}C n_\mathrm{h},
\end{equation}
where $m_\mathrm{spin}$ and $m_T^\theta$ are the spin-moment and magnetic dipole moment, respectively. The quantities, $p$, $q$, and $r$ are the XMCD integrations for the $L_3$-edge, $L_3 + L_2$-edges, and the integral of XAS which are defined as the following equations: $p = \int_{L_3} (\mu^- - \mu^+) dE$, $q = \int_{L_3+L_2} (\mu^- - \mu^+) dE$, and $r = \int_{L_3+L_2} (\mu^- + \mu^+ - \mathrm{b.g}) dE$ (b.g is a background spectrum showing a step-function shape).
A correction factor ($C$) is included in equation \eqref{eq:mspin} to take into account the so-called \textit{jj} mixing effect~\cite{Teramura1996,Durr1997,Goering2005}. For $m_\mathrm{spin}^\mathrm{eff}$ of Mn and Co, the $C$ of 1.5 and 1.1 were assumed, respectively~\cite{Teramura1996,Durr1997,Nagai2018}. The numbers of 3d holes ($n_\mathrm{h}$) were assumed to be 4.44 $\pm$ 0.08 and 2.29 $\pm$ 0.02 for the Mn and Co atoms, respectively, based on the theoretical values for Mn$_2$CoGa obtained from the first principles calculations in section \ref{sec:calc}.
Here, the values of $n_h$ are averaged over four cases of Mn$_2$CoGa with X$_\mathrm{A}$ and L2$_{1\mathrm{b}}$ phases and $c/a$ ratios of 1.00 and 1.04. The values of error are the standard deviation of all cases including possible inequivalent atomic sites for Mn atoms which are described in detail later.
In a strict manner,  the inequivalent sites should be deconvoluted from the XAS$/$XMCD spectra, and the equations \eqref{eq:morb} and \eqref{eq:mspin} are applied to each site, however, the present $n_\mathrm{h}$ for the Mn sites showing similar values enable us to evaluate average values using the equations \eqref{eq:morb} and \eqref{eq:mspin} as an approximation~\cite{Takata2018}.
The values of $m_\mathrm{spin}^\mathrm{eff}$, $m_\mathrm{orb}^\theta$ and $m_\mathrm{orb} / m_\mathrm{spin}^\mathrm{eff}$ are summarized in figure~\ref{fig:sumrule}. Corresponding total magnetic moment was calculated to be $1.18 \pm 0.03~\mu_\mathrm{B}/\mathrm{f.u.}$ using $m_\mathrm{spin}^\mathrm{eff}$ and $m_\mathrm{orb}^\theta$ at $\theta =$ 54.7\degree, and the film composition. The total magnetic moment by the sum rule is of the same order as the $\mu_\mathrm{s}$ of 1.27 $\mu_\mathrm{B}/\mathrm{f.u.}$ measured by VSM. Concerning the angular dependence, no $\theta$ dependence was found either for $m_\mathrm{spin}^\mathrm{eff}$ or $m_\mathrm{orb}^\theta$.
Here, the $\theta$ of 54.7\degree\ is a magic angle for which $m_T^\theta=0$ by applying the approximate symmetry relation of $2m_T^{90\degree} + m_T^{0\degree} = 0$, where $m_T^\theta = m_T^{90\degree}\sin^2\theta + m_T^{0\degree}\cos^2\theta$ for 3d metals~\cite{Stohr1995,Weller1995}. Thus, the angular independent $m_\mathrm{spin}^\mathrm{eff}$ values suggest that the anisotropy of the magnetic dipole moment ($\Delta m_\mathrm{T} = m_T^{0\degree} - m_T^{90\degree} = \frac{3}{2}m_T^{0\degree}$) is less than 0.036 $\mu_\mathrm{B}$/atom which is in an error-bar range in the present sample.
The angular independent $m_\mathrm{orb}^\theta$ values provide information for the discussion of origin of PMA in series B:
The magnetocrystalline anisotropy energy ($E_\mathrm{MCA}$) has been theoretically evaluated using expressions based on the second-order perturbation theory in terms of spin-orbit interactions~\cite{Bruno1989,Laan1998,Kota2014}. 
Note that $E_\mathrm{MCA}$ is defined as the energy difference when the magnetization points to the perpendicular and in-plane direction: $E_\mathrm{MCA} = E^{\theta=90} - E^{\theta=0}$, \textit{i.e.}, the positive sign of $E_\mathrm{MCA}$ indicates PMA in this definition. Within a formulation $E_\mathrm{MCA}$ can be expressed as follows~\cite{Bruno1989,Laan1998}:
\begin{equation}
\label{eq:Bruno}
E_\mathrm{MCA} \simeq \frac{\xi}{4} \Delta m_\mathrm{orb} - \frac{21}{2}\frac{\xi^2}{E_\mathrm{ex}}\Delta m_\mathrm{T},
\end{equation}
where $\xi$, $E_\mathrm{ex}$, and $\Delta m_\mathrm{orb}$ are the spin--orbit coupling constant, exchange splitting of 3d bands, and the anisotropy of orbital magnetic moment defined as $\Delta m_\mathrm{orb} = m_\mathrm{orb}^{\theta = 0} - m_\mathrm{orb}^{\theta = 90\degree}$, respectively, for which the first term on the right side corresponds to the contribution to the energy difference from the spin-conservation terms in virtual excitations, and the second term corresponds to that from the spin-flip term.
The experimentally observed angular independent $m_\mathrm{orb}^\theta$ implies small contribution of the spin-conservation terms to $K_\mathrm{u}$, and the other contribution of the spin-flip terms discussed in refs. \onlinecite{Laan1998,Kota2014} is more important.
Theoretical values of $K_\mathrm{u}$ and $m_\mathrm{orb}$ as well as the origin of PMA with respect to the spin-flip term are to be discussed further in section~\ref{sec:calc}.

The sub-peaks and shoulder structures in XAS and XMCD spectra provide additional information on the chemical order of the atoms in Heusler structures~\cite{Miyamoto2003,Telling2006,Kubota2009a,Klaer2011,Meinert2011,Ouardi2015,Karel2015,Nagai2018,Yoshikawa2020}.
Note that a possibility of oxidization is excluded, because the shapes of XAS and XMCD spectra are completely different from those reported for materials containing Mn- and/or Co-oxides~\cite{Burnus2008,Jung2009}. In addition, shoulders and sub-peaks are clearly observed at 300 K for MCD spectra shown in Figs.~\ref{fig:MCD}(c) and~\ref{fig:MCD}(d), for which the measurement temperature is higher than the magnetic transition temperatures for most of Mn-/Co-oxide materials. 
\begin{figure}
\includegraphics[clip,scale=1.0]{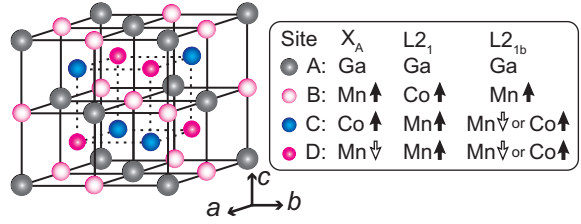}
\caption{
Schematic illustration of the Heusler structure, and possible atomic sites for Mn$_2$CoGa. Relative directions of magnetic moments for each atom theoretically reported in ref. \onlinecite{Umetsu2018} are described using black and white arrows beside the name of atoms.
}
\label{fig:Heusler}
\end{figure}
Figure~\ref{fig:Heusler} shows a schematic illustration of a crystal structure for Heusler alloys and the notations of the atomic sites for the inverse-Heusler structure (X$_\mathrm{A}$ phase), the full-Heusler structure (L2$_1$ phase), and an L2$_\mathrm{1b}$ phase.
In addition to these three phases, fully-disordered D0$_3$ phase was considered as possible chemical phases in previous studies on bulk samples~\cite{Minakuchi2015,Umetsu2018}.
According to total energy calculations in ref. \onlinecite{Umetsu2018}, the X$_\mathrm{A}$ phase was the most stable at the ground state, and relatively small energy difference of 0.136 eV/f.u. was reported for the L2$_{1\mathrm{b}}$ phase. The energy differences with respect to the X$_\mathrm{A}$ phase were much higher for the L2$_1$ and D0$_3$ phases than that for the L2$_{1\mathrm{b}}$ phase.
Concerning the magnitudes of magnetic moment for each atom, the values were similar in X$_\mathrm{A}$ and L2$_{1\mathrm{b}}$ phases to each other and ferrimagnetic coupling was reported as schematically shown in Fig.~\ref{fig:Heusler}, which resulted in a total moment of about 2.0 $\mu_\mathrm{B}$/f.u. for both. On the other hand, those were relatively large (small) for the L2$_1$ (D0$_3$) phase showing ferromagnetic (ferrimagnetic) coupling with a total moment of about 7.7 (1.0) $\mu_\mathrm{B}$/f.u.
As it was discussed on Fig.~\ref{fig:mag} in Sec.~\ref{sec:mag}, the total magnetic moments of the present samples are of the order of the reported values for the X$_\mathrm{A}$ and L2$_{1\mathrm{b}}$ phases, and considering the calculated energy difference mentioned above, the L2$_1$ and D0$_3$ phases are excluded in the following discussion.
Note that the L2$_{1\mathrm{b}}$ phase was experimentally proposed for the previous bulk Mn$_2$CoGa samples based on high-angle annular dark field scanning electron transmission microscope images~\cite{Minakuchi2015} and neutron diffraction experiments~\cite{Umetsu2018}. On the other hand, the X$_\mathrm{A}$ phase was proposed in another previous study on an epitaxially grown Mn$_2$CoGa film based on XAS and XMCD results~\cite{Meinert2011}.
The shapes of XAS and XMCD spectra of Mn$_2$CoGa films were previously discussed based on the first-principles calculation for the X$_\mathrm{A}$ phase. In the previous studies~\cite{Meinert2011,Ouardi2015}, several shoulders, sub-peaks, and dip structures were mentioned. In the present results, similar structures are observed, \textit{i.e.}, broad shoulders in the XAS, and sub-peaks and dip structures in the XMCD spectra are observed, which are marked by solid and broken lines in Fig. \ref{fig:MCD}.
The marked structures in the XMCD spectra can also be understood as a sum of two spectra~\cite{Klaer2011,Klaer2011a,Meinert2011,Ouardi2015,Karel2015} which are, \textit{e.g.}, MVA~\cite{Nagai2018} and CMG~\cite{Yoshikawa2020} for the markers of broken and solid lines, respectively.
Here, MVA is a typical example of a material with Mn atom(s) at D(C)-site, and CMG is another example with a Mn atom at B-site in Fig.~\ref{fig:Heusler}. 
The broad feature is mainly from the Mn atom(s) D(C)-site showing relatively itinerant nature of electrons~\cite{Nagai2018}, on the other hand, the Mn atom at the B-site exhibits relatively localized nature showing remarkable shoulders in XAS~\cite{Miyamoto2003,Telling2006,Yoshikawa2020}.
Concerning the XMCD spectra for Co, shoulders are also observed around the post-absorption edges with the solid markers in Fig.~\ref{fig:MCD}(d) which also includes a reference spectrum of CMG. The origin of the shoulders is attributed to the 2p $\rightarrow$ 3d $e_g$ (or $e_u$) transition~\cite{Yoshikawa2020}.

\section{First-principles calculation}
\label{sec:calc}

First-principles calculation of electronic structure and magnetic properties of bulk Mn$_2$CoGa with the X$_\mathrm{A}$ and L2$_\mathrm{1b}$ phases were performed by employing the tight-binding linear muffin-tin orbital method under the local spin-density functional approximation~\cite{Turek1996}. The coherent potential approximation was adopted to treat the partial disorder between Mn and Co in C- and D-sublattices of the L2$_\mathrm{1b}$ phase (see Fig.~\ref{fig:Heusler}). The lattice constant $a_0 = 0.5879$~nm~\cite{Umetsu2018} was adopted. We have confirmed that the electronic structure is half-metallic in both phases and the calculated magnetization is close to integer 2~$\mu_\mathrm{B}$/f.u., which is consistent with a previous calculation by Umetsu, Tsujikawa \textit{et al}~\cite{Umetsu2018}. For the calculation of $K_\mathrm{u}$ and $m_\mathrm{orb}$, spin--orbit interaction was taken into account~\cite{Turek2008}. The $K_\mathrm{u}$ value was evaluated from the energy difference between when the magnetization aligned along the $c$- and $a$-axis directions, \textit{i.e.}, $K_\mathrm{u} = (E_a - E_c)/V$~\cite{Kota2012a,Kota2012b}, within the magnetic force theorem ($V$: volume of a unit cell). The number of k-point sampling was set to 55,440 in the full-Brillouin zone. 

\begin{figure}[t]
\includegraphics[width=0.4\linewidth]{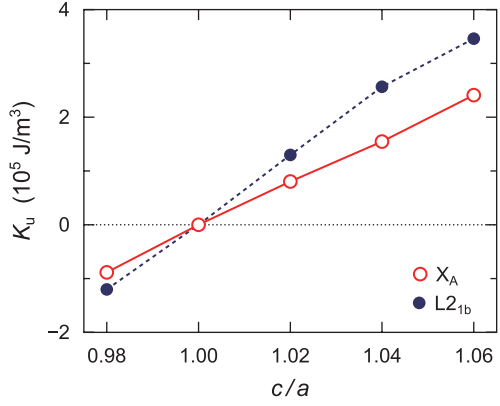} 
\caption{
Axial ratio $c/a$ dependence of uniaxial magnetic anisotropy constant $K_\mathrm{u}$ of Mn$_2$CoGa with the X$_\mathrm{A}$ and L2$_\mathrm{1b}$ phases obtained from first-principles calculation.
}
\label{fig:calc1}
\end{figure}

\begin{figure}[t]
\includegraphics[width=0.5\linewidth]{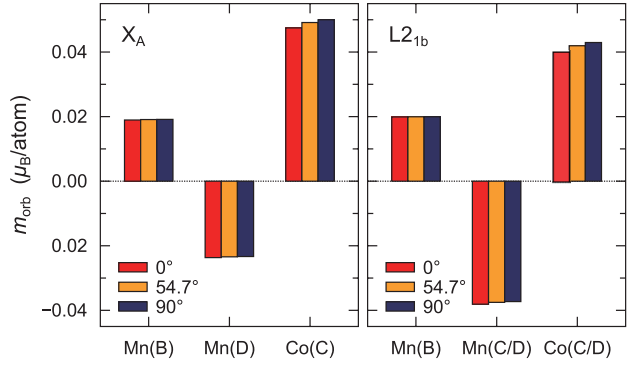}
\caption{Calculated orbital magnetic moment $m_\mathrm{orb}$ of Mn and Co atom in Mn$_2$CoGa with the X$_\mathrm{A}$ and L2$_\mathrm{1b}$ phases. The angle of 0$^\circ$, 54.7$^\circ$, and 90$^\circ$ means the relative angle between magnetization and crystal $c$-axis.}
\label{fig:calc2}
\end{figure}

Figure~\ref{fig:calc1} shows the calculated result of $K_\mathrm{u}$ of Mn$_2$CoGa in the X$_\mathrm{A}$ and L2$_\mathrm{1b}$ phases as a function of the axial ratio $c/a$, where $c/a$ was changed under the assumption of constant volume deformation. PMA appears for $c/a>1$ in both phases. The $K_\mathrm{u}$ values at $c/a = 1.04$ in the X$_\mathrm{A}$ and L2$_\mathrm{1b}$ phases are $1.5 \times 10^5$~J/m$^3$ (0.049~meV/f.u.) and $2.6 \times 10^5$~J/m$^3$ (0.081~meV/f.u.), respectively, which are quantitatively consistent values with the obtained experimental result shown in Fig.~\ref{fig:Ku}. Figure~\ref{fig:calc2} shows the $m_\mathrm{orb}$ of Mn and Co atoms when the relative angle between magnetization and crystal $c$-axis is 0\degree, 54.7\degree, and 90\degree~for $c/a = 1.04$. Note that the positive (negative) sign of $m_\mathrm{orb}$ means that the direction of orbital magnetic moment and overall magnetization are the same (opposite). Thus, the directions of spin and orbital magnetic moments are the same each other in Mn at D-sublattice with X$_\mathrm{A}$ phase and in Mn at C/D-sublattice with the L2$_\mathrm{1b}$ phase, because the spin magnetic moment is arranged in the opposite direction with the magnetization. This result is in agreement with a previous first-principles calculation~\cite{Meinert2011}. In Fig.~\ref{fig:calc2}, we find that the $\Delta m_\mathrm{orb}$ is of the order of 0.001~$\mu_\mathrm{B}$/atom in Co and even smaller in Mn. Based on Eq.~\eqref{eq:Bruno}, therefore, the orbital moment anisotropy cannot explain the magnetocrystalline anisotropy observed in our experiment. This fact implies that the PMA in Mn$_2$CoGa is mainly induced by the hybridization of $\uparrow$-spin and $\downarrow$-spin states via spin--orbit coupling, which is similar to the discussion of magnetocrystalline anisotropy in previous studies~\cite{Miwa2017,Ikeda2017,Okabayashi2020}.\\

\begin{figure*}[t]
\includegraphics[width=0.9\linewidth]{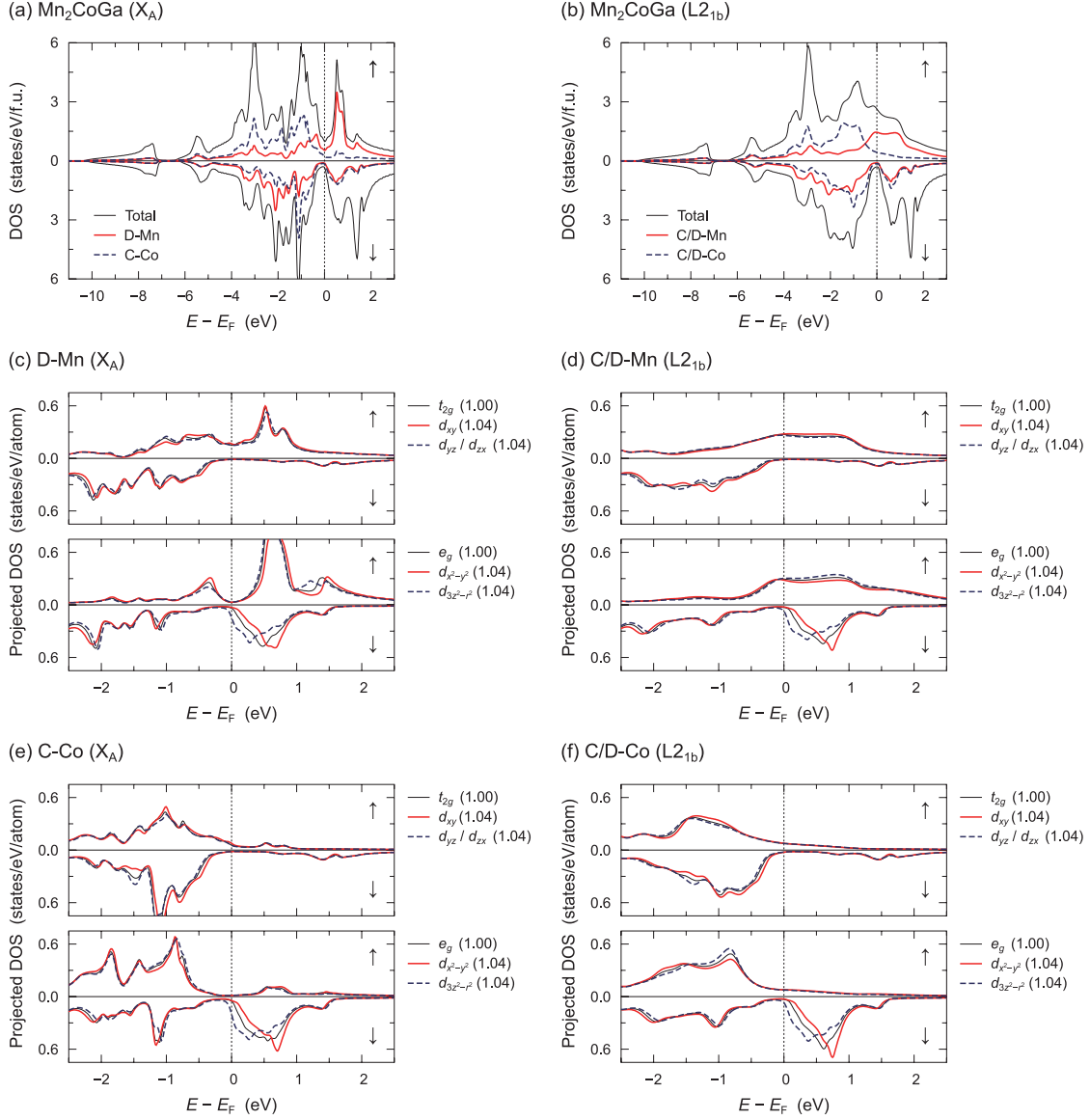}
\caption{(a), (b) Total and local DOS of Mn and Co at C- and/or D-sublattice in Mn$_2$CoGa with the X$_\mathrm{A}$ and L2$_\mathrm{1b}$ phases for $c/a$ = 1.00. (c)--(f) Projected DOS into $t_{2g}$-based $d_{xy}$, $d_{yz}$, and $d_{zx}$ orbitals, and into $e_g$-based $d_{x^2-y^2}$ and $d_{3z^2-r^2}$ orbitals for $c/a$ = 1.00 and 1.04. The values in parenthesis denote $c/a$. The upper and lower panels of each figure present the DOS of $\uparrow$- and $\downarrow$-spin states, respectively. }
\label{fig:calc3}
\end{figure*}
For a detailed analysis of magnetocrystalline anisotropy, we introduce the expression of $K_\mathrm{u}$ based on the second-order perturbation theory in terms of the spin--orbit coupling,
\begin{equation}
K_\mathrm{u} = \frac{\xi^2}{4V} \sum_{\mathrm{o}\sigma} \sum_{\mathrm{u}\sigma'} \frac{2\delta_{\sigma,\sigma'}-1}{\varepsilon_{\mathrm{u}\sigma'}-\varepsilon_{\mathrm{o}\sigma}} \big[ \vert \langle \mathrm{u}^{\sigma'} \vert \ell_z \vert \mathrm{o}^\sigma \rangle \vert^2  -  \vert \langle \mathrm{u}^{\sigma'} \vert \ell_x \vert \mathrm{o}^\sigma \rangle \vert^2 \big],
\label{eq:Ku}
\end{equation}
where $\varepsilon_{\mathrm{o} \sigma}$ and $\vert \mathrm{o}^\sigma \rangle$ ($\varepsilon_{\mathrm{u} \sigma}$ and $\vert \mathrm{u}^\sigma \rangle$) denote the eigenvalue and eigenvector of the nonpertubative occupied (unoccupied) states, respectively ($\sigma = \ \uparrow \mathrm{or} \downarrow$). In addition, $\ell_z$ ($\ell_x$) is the $z$ ($x$) component of the orbital angular momentum operator. From the sublattice decomposition analysis of magnetocrystalline anisotropy energy~\cite{Solovyev1995,Ke2019}, we confirmed that Mn and Co in C- and/or D-sublattice mainly yields the PMA in tetragonally distorted Mn$_2$CoGa for $c/a > 1$ in both of the X$_\mathrm{A}$ and L2$_\mathrm{1b}$ phases~\footnote{
The estimated anisotropy energy of Mn and Co in C- and/or D-sublattice using the formulation in Ref.~\cite{Solovyev1995} is on the order of 0.01~meV with PMA, and further that of B-Mn is one order of magnitude smaller. Therefore, the contribution of B-Mn is relatively small compared to the contribution of the other Mn and Co.}.
In figure~\ref{fig:calc3}, we depict the density of states (DOS) of X$_\mathrm{A}$- and L2$_\mathrm{1b}$-Mn$_2$CoGa. Figs~\ref{fig:calc3}(a) and \ref{fig:calc3}(b) show the total DOS and the local DOS of Mn and Co at C- and/or D-sublattice for the cubic ($c/a=1.00$) case. Furthermore, Figs.~\ref{fig:calc3}(c)--\ref{fig:calc3}(f) show the DOS projected into the $d$-orbitals ($d_{xy}$, $d_{yz}$, $d_{zx}$, $d_{x^2-y^2}$, $d_{3z^2-r^2}$) near the Fermi energy ($E_\mathrm{F}$) for the cubic and tetragonal ($c/a=1.04$) cases. The degeneracy in $t_{2g}$ ($d_{xy}$, $d_{yz}$, $d_{zx}$) based states and that in $e_g$ ($d_{x^2-y^2}$, $d_{3z^2-r^2}$) based states are removed by the tetragonal distortion. In particular, we can see a significant DOS splitting in $d_{x^2-y^2}$ and $d_{3z^2-r^2}$ states, because $d_{3z^2-r^2}$ ($d_{x^2-y^2}$) orbital spreads to the parallel (perpendicular) direction to the crystal $c$-axis.

Let us discuss a possible origin for PMA in tetragonally distorted Mn$_2$CoGa qualitatively, using the expression of Eq.~\eqref{eq:Ku} and the projected DOS shown in Figs.~\ref{fig:calc3}(c)--\ref{fig:calc3}(f). With reference to previous studies~\cite{Kota2014,Okabayashi2020,Masuda2021}, we assumed $\varepsilon_{\mathrm{u}\sigma} - \varepsilon_{\mathrm{o}\sigma'}$ in Eq.~\eqref{eq:Ku} as a constant value which is independent of wave-number $k$ for simplicity~\footnote{The $k$-resolved analysis is desirable for more detailed investigation of $K_\mathrm{u}$ based on the second order perturbation theory. However, this treatment is considered difficult to apply to bulk systems including several species of atoms and randomness, compared to slab geometry systems such as ferromagnetic/nonmagnetic layer junctions.~\cite{Masuda2017,Masuda2018,Kwon2019,Jiang2020} Thus, we give a simplified analysis for discussion.}.
In Figs.~\ref{fig:calc3}(c)--\ref{fig:calc3}(f), the peaks of $d_{x^2-y^2}^\downarrow$ and $d_{3z^2-r^2}^\downarrow$ states are located just above $E_\mathrm{F}$, and also the $d_{yz}^\uparrow$ state lies below $E_\mathrm{F}$, in Mn and Co at C- and/or D-sublattice with both of the X$_\mathrm{A}$ and L2$_\mathrm{1b}$  phases. These states hybridize each other via spin--orbit coupling, which results in non-zero matrix element of $\langle d_{yz} \vert \ell_x \vert d_{x^2-y^2} \rangle$ and $\langle d_{yz} \vert \ell_x \vert d_{3z^2-r^2} \rangle$. In Eq.~\eqref{eq:Ku}, the mixing of occupied $d_{yz}^\uparrow$ state and unoccupied $d_{x^2-y^2}^\downarrow$ state, and the mixing of occupied $d_{yz}^\uparrow$ state and unoccupied $d_{3z^2-r^2}^\downarrow$ state make for the positive $K_\mathrm{u}$ value, \textit{i.e.}, promoting PMA. Note here that $\delta_{\uparrow\downarrow}=0$ in eq. \eqref{eq:Ku}. In Figs.~\ref{fig:calc3}(c)--\ref{fig:calc3}(f), the peak of the unoccupied $d_{3z^2-r^2}^\downarrow$ state of Mn and Co shifts to lower energy side with increasing $c/a$; this enhances the positive contribution of $K_\mathrm{u}$, because the energy difference from the occupied $d_{yz}^\uparrow$ state, $\varepsilon_{\mathrm{u}\downarrow}-\varepsilon_{\mathrm{o}\uparrow}$, becomes smaller, in Eq.~\eqref{eq:Ku}. In contrast, the peak of $d_{x^2-y^2}^\downarrow$ shifts to the higher energy side, reducing the positive contribution. We notice that there is a tradeoff relation enhancing and reducing the positive $K_\mathrm{u}$ with increasing $c/a$; however, the former contribution is three times as large as the latter, because of $\vert \langle d_{yz} \vert \ell_x \vert d_{3z^2-r^2} \rangle \vert^2 = 3$ and $\vert \langle d_{yz} \vert \ell_x \vert d_{x^2-y^2} \rangle \vert^2 = 1$. Therefore, the PMA is induced by increasing the positive $K_\mathrm{u}$ value in the tetragonally distorted MCG films for $c/a > 1$ through the spin-flip term in Eq.~\eqref{eq:Ku}, and a qualitative mechanism for the PMA is almost the same regardless of whether crystal structure is the X$_\mathrm{A}$ or L2$_\mathrm{1b}$ phases. On the other hand, the in-plane magnetic anisotropy arises for $c/a < 1$, as shown in Fig.~\ref{fig:calc1}, based on the opposite scenario of the above.

\section{Summary}
Crystal structure and magnetic properties of MCG films on Cr or Ag buffer were investigated experimentally and theoretically. Epitaxial growth of all samples was confirmed using XRD profiles, and the different growth modes were suggested depending on the buffer materials, which possibly caused the different $c/a$ ratios among the samples. Tetragonal distortion of the MCG layers was found for the Cr buffer samples up to the layer thickness of 30 nm, whereas Ag buffer samples exhibited cubic structure which was stable in bulk Mn$_2$CoGa samples in literature. From magnetization curves measurements, perpendicular magnetization was found in all the Cr buffer samples; on the other hand, all Ag buffer samples exhibited in-plane magnetization. The $c/a$ ratio dependence of $K_\mathrm{u}$ clearly exhibited positive correlation, and the maximum value of $K_\mathrm{u}$ was 1.6 $\times$ 10$^5$ J/m$^3$ at room temperature for the 5-nm-thick MCG film on the Cr buffer. 
Magnetic domain pattern observations using the magneto-optical Kerr microscopy revealed changes of the domain structures depending on the magnitude of $K_\mathrm{u}$.
XAS and XMCD spectra were also acquired to elucidate the origin of magnetocrystalline anisotropy in MCG, and unraveled the chemical order of L2$_{1\mathrm{b}}$ or X$_\mathrm{A}$ phase. In addition, the angular dependence of XMCD spectra revealed that the change of the $m_\mathrm{orb}$ was negligibly small suggesting a negligible contribution of the spin conserved term, so called Bruno's term, to $K_\mathrm{u}$.
Theoretical values of $K_\mathrm{u}$ were evaluated by means of the first principles electronic structure calculation. The positive correlation was also confirmed between the $c/a$ ratios and the values of  $K_\mathrm{u}$ in the calculation. In addition, the calculated angular dependence of $m_\mathrm{orb}$ was negligibly small. Those results are consistent with the experimental results. Based on the analysis of the second-order perturbation theory, the origin of the magnetocrystalline anisotropy is proposed that the hybridization between two crystal-field split states with opposite spins, \textit{i.e.}, the spin-flip term near the Fermi level, promotes positive contribution to $K_\mathrm{u}$.

\begin{acknowledgments}
The XAS and XMCD experiments were  performed under the Shared Use Program of JAEA Facilities (2018B-E24, and 2019B-E17) with the approval of the Nanotechnology Platform project supported by the Ministry of Education, Culture, Sports, Science and Technology (A-18-AE-0042, and A-19-AE-0037). The synchrotron radiation experiments were performed at the JAEA beam line BL23SU in SPring-8 (2018B3842, and 2019B3844).
TK, DT and MM would like to thank Mr. Issei Narita for technical support.
SB, SM thank department of atomic energy (DAE), Government of India for financial support for the Kerr microscopy facility.
This work was partially supported by KAKENHI (JP20K05296) from JSPS, and by a cooperative research program (No.20G0414) of the CRDAM-IMR, Tohoku Univ.
\end{acknowledgments}


\section*{Conflict of Interest}
The authors have no conflicts to disclose.

\section*{Data Availability Statements}
The data that support the findings of this study are available within its supplementary material, and from the corresponding author upon reasonable request.

\bibliographystyle{aipnum4-2}
\bibliography{
d:/Lab/Manuscripts/BibTex/Heusler_etc,
d:/Lab/Manuscripts/BibTex/Mn-alloys,
d:/Lab/Manuscripts/BibTex/Oxides,
d:/Lab/Manuscripts/BibTex/XMCD,
d:/Lab/Manuscripts/BibTex/Surface,
d:/Lab/Manuscripts/BibTex/Theory
}

\end{document}